\documentstyle[12pt]{article}
\input epsf

\renewcommand{\Large}{\large}
\catcode `@=11 \@addtoreset{equation}{section} \catcode `@=12

\begin{document}

\def\gap#1{\vspace{#1ex}}
\def\be{\begin{equation}}
\def\ee{\end{equation}}
\def\ba{\begin{array}{l}}
\def\ea{\end{array}}
\def\bea{\begin{eqnarray}}
\def\eea{\end{eqnarray}}
\def\eq#1{(\ref{#1})}
\def\del{\partial}
\def\A{{\cal A}}
\def\sh{{\rm sinh}}
\def\ch{{\rm cosh}}
\def\z{{\bar z}}
\def\V{{\tt V}}	
\def\M{{\cal M}}
\def\BE{{\rm b.e.}}
\def\G{\Gamma}
\def\q{{\tt q}}
\def\no#1{{\tt   hep-th #1}}

\renewcommand\arraystretch{1.5}

\begin{flushright}
TIFR-TH-96/36 \\
CERN-Th/96-294 \\
hep-th/9610120
\end{flushright}
\begin{center}
\gap{3}
{\Large\bf 
Observability of quantum  state of black hole
}\\
\gap{10}
Justin R. David$^*$, Avinash Dhar$^*$, Gautam Mandal$^*$\\
Tata Institute of Fundamental Research \\
Homi Bhabha Road, Mumbai 400 005, INDIA \\
\gap{2}
and\\
\gap{2}
Spenta R. Wadia$^{*\, \dagger}$ \\
Theory Division, CERN, CH-1211, Geneva 23, Switzerland \\
\gap{15}
\bf ABSTRACT\\
\end{center}
\gap{2}
We analyze terms subleading to Rutherford in the $S$-matrix between
black hole and probes of successively high energies.  We show that by
an appropriate choice of the probe one can read off the quantum state
of the black hole from the S-matrix, staying asymptotically far from
the BH all the time. We interpret the scattering experiment as
scattering off classical stringy backgrounds which explicitly depend
on the internal quantum numbers of the black hole.
\vfill
\hrule
\gap{-1}
\begin{flushleft}
$^*$ {\tenrm e-mail: justin, adhar, mandal,
wadia@theory.tifr.res.in}\\ 
$^\dagger$ {\tenrm On leave from Tata
Institute of Fundamental Research, India.}
\end{flushleft}
\eject

\setcounter{section}{-1}

\section{Introduction} 

One of the most mysterious aspects of black hole physics is the
no-hair theorem. In its classic form it appears to declare
unobservable any attribute other than the mass $M$, charge $Q$ and
angular momentum $J$ of a black hole
\cite{classic}. The mystery deepens with Hawking's
discovery that even after quantum effects are switched on, the
radiation emitted is apparently thermal, with a temperture $T(M, Q,
J)$, thus limiting the observable information once again to those
three quantities. In the context of a gravitational collapse, this
would appear to suggest (to the radically-inclined) that the process
of collapse reduces any arbitrary configuration of collapsing matter
to a {\em unique} quantum state characterized by the above three
quantities. While the violation of various global conservation laws
implied by such a scenario is not entirely unthinkable, many distinct
initial states evolving to the same final state is inconsistent with
unitarity: a basic tenet of quantum mechanics. Furthermore, it rules
out any thermodynamic understanding of the Hawking-Beckenstein entropy
as the logarithm of the number of states.

One possible way out of this would be to say that there is no unique
final state, but so far as the external world is concerned, all those
states are {\em indistinguishable}. Although such a position avoids
the two objections mentioned in the last paragraph, it seems to allow,
in a sense, unobservable {\em observables}.\footnote{This
is related to the more general  issue of complementarity of 
descriptions of physical phenomena in the context of black holes.}

In this paper we study the issue of {\em observability} of the
``internal'' states of the black holes in the context of black hole
models in string theory. We begin the discussion by considering the
electrically charged black holes of \cite{sen}(Sec. 1). The states
here are given explicitly by conformal field theory vertex
operators. Various states differ in the choice of internal
polarizations in the compact directions. In \cite{manwad} the problem
of scattering of probes off such black holes was considered. In the
limit of a black hole of large mass and low energy of probe
(Rutherford limit) it was found that leading term in the scattering
could be understood as scattering off a black hole metric which in
particular did not depend on the internal polarization of the black
hole. Already in \cite{manwad} it was found that this no-hair property
did not persist in higher order terms in the scattering matrix (beyond
Rutherford) and that these showed dependence on the internal
polarization tensors of the black hole. In this paper we take a more
detailed look at the post-Rutherford terms.  We find that they involve
non-trivial entanglement between polarization tensors of the BH and
polarization tensors (and charges) of the probe. Indeed, not only can
we get {\em some} information about the internal polarization of the
black hole from the S-matrix, by using appropriate probes, we can
actually completely determine the state of the black hole. In other
words, measurements from far away can uniquely determine the quantum
state of the black hole. Thus, the black hole states are no different
from ordinary string states in this regard.  Recently Susskind et al
\cite{sus} have argued that {\em any} (sufficiently massive) string
state is a black hole after all. The observation that we have made
above is consistent with this proposal. 

The fact that the the scattering matrix contains more information
about the quantum state of the black hole, beyond that allowed by the
classical no-hair theorems, and the fact that the scattering matrix
has the interpretation of scattering of probe particles off
backgrounds of various string modes around the BH (Sec. 2), naturally
leads us to ask the question: How do these backgrounds manage to carry
the additional information? Larsen and Wilczek \cite{larwil} have
argued that the usual treatment of no-hair theorem is rather
restrictive and that within a higher dimensional field theory
Kaluza-Klein framework there are an infinite number of distinct
classical solutions, corresonding to the same overall mass and
charges, but differing in the background values of the Kaluza-Klein
modes.  We show that (Sec. 2) our scattering results imply that the
elementary BPS state not only gives rise to backgrounds of metric,
dilaton, etc. but also to backgrounds of an infinite number of higher
string modes. Although the metric and dilaton backgrounds do not
depend on the internal polarization tensor (at leading order), the
massive string modes explicitly depend on them, e.g.
\be
M_{ijkl}(\rho) = (1/m) e^{- \sqrt{2} \rho}/\rho \; [\zeta_{R,i}
\zeta_{R,j} +  Q_{R,i}Q_{R,j}]
[\zeta_{L, k}\zeta_{L,l} + 1/2 Q_{L,k}Q_{L,l}] + o(g_{st})
\label{0.1}
\ee
(see equation \eq{2.13} in Sec. 2). It turns out that by measuring a
sufficient number of these backgrounds we can actually determine the
microstate of the black hole entirely. Inelastic amplitudes involving 
Kaluza-Klein charge exchange between the probe particles and
the black hole also contain terms which show dependence on
the internal polarization of the black hole, and, therefore, 
can be used to get information on the microstate 
of the black hole.  These amplitudes seem to have a close connection
with the observations made in \cite{larwil}.

We believe that the scenario that we have presented above suggests
that string theoretic models of black holes come packaged with an
infinite number of higher mass string backgrounds (in addition to
metric, dilaton and moduli) which contain information about the
detailed state of the black hole. In Sec. 3 we make some remarks about
hair on D-brane models of black holes. In Sec. 4 we present a
summary and outlook. 

\gap{3}

\section{Hair from $S$-matrix}

In this section we consider the electrically charged black holes of
\cite{sen} and show that by scattering suitable probes off such a
black hole it is possible to get information about (and, in fact, {\em
determine}) its  detailed quantum state.

We work in heterotic string compactified on $T^6$. Our notations are
as follows (for details, see \cite{manwad}). The bosonic coordiantes
are $ x^\mu, \mu = 0,1,2,3, \, x^i_R, i= 1, \ldots, 6, \, x^i_L, i =
1, 2, \ldots, 22$. $x^i_{R,L}$ are holomorphic and antiholomorphic
respectively. World sheet fermions are $\psi^\mu (z)$ and
$\psi^i_R(z)$. Here $R,L$ stand for right and left movers (analytic
and antianalytic repectively in our convention). For a generic torus
$T^6$, the gauge group is abelian: $U(1)^{28}$, arising from 6
right-moving and 22 left-moving currents. We will denote the
corresponding charges as $\vec Q_R$ and $\vec Q_L$ resp. For BPS
states the mass $m$ satisfies the condition
\be
m^2 =  Q_R^2 = Q_L^2 + 2 (N_L -1 )
\label{1.1}
\ee
where $N_L$ is the oscillator level in the left-moving (antianalytic)
sector.  The black holes of \cite{sen} are BPS states represented by
the vertex operators of the form
\be
\ba
\V_B(\zeta_R; \zeta_L;k; z,\z) = 
V_B(\zeta_R, k, z) \bar V_B(\zeta_L, \z) \exp[iQ_R.x_R+ iQ_L.x_L +
ik.x(z,\z)]\\ 
V_B(\zeta, k,z) = \zeta_R. \psi_R (z) \, e^{-\phi(z)} \\
\bar V_B(\zeta_L, \z) = \zeta_{L,i_1 i_2 \ldots i_r} 
\del_\z^{n_1}  x^{i_1}_L 
\del_\z^{n_2}  x^{i_2}_L \ldots
\del_\z^{n_r}  x^{i_r}_L 
\\
\ea
\label{1.2}
\ee
In the above $k^2 = - m^2$ and 
\be
N_L = \sum_i n_i
\label{1.3}
\ee
Different black holes with the same mass $m$ and charges $\vec
Q_{R,L}$ differ in the choice of the internal polarization tensor
$\zeta_L$ (they differ in $\zeta_R$ also, but for large $N_L$ the main
degeneracy comes from varying $\zeta_L$'s).  For simplicity we have
here chosen the polarization tensors $\zeta_{L,R}$ entirely in the
compact directions.

In the following we will use various probes to extract information
about the state of the black hole, or in other words about the
polrization $\zeta_L$ (and $\zeta_R$). We discuss the various choices
of probes in turn:

\gap{2}

(a) Massless probes:

\gap{2}

These are given by vertex operators

\be
\ba
\V_P(\eta_R; \eta_L; k; z,\z) = 
V_P(\eta_R,k, z) \bar V_P(\eta_L, \z) \exp[ik.x(z,\z)]\\
V_P(\eta_R ,k, z) =  \eta_{R,M}(\del_z x^M + i k_\mu\psi^\mu 
\,\psi^M) \\
\bar V_P(\eta_L, \z) = \eta_{L,N}
\del_\z  x^N 
\\
\ea
\label{1.4}
\ee

The case when the polarization vectors $\eta_{R,L}$ have components
only in the compact directions corresponds to the moduli fields as
probe. The four point amplitude \cite{manwad} describing the
scattering of these probes off the black hole is as follows.

\gap{2}

\underbar{$N_L=1$ black holes}

\gap{2}

For $N_L =1$ black holes the amplitude is%
\be
\ba
\M(1,2,3,4) = A_1(s,t,u) \times \\
\,~\big( \zeta_R. \zeta'_R \, \eta_R. \eta'_R
+ [ {t \over s - m^2} \zeta_R. 
\eta_R \, \zeta'_R. \eta'_R + \BE ]  - {2 t \over
(s-m^2)(u - m^2)} \eta_R.\eta'_R \, \zeta.Q_R \, \zeta'.Q_R
\big)  \times
\\
\big ( \zeta_L. \zeta'_L \, \eta_L. \eta'_L
 + [ {(t + 2)t \over(2 + s- m^2) (s- m^2)} \zeta_L. 
\eta_L \, \zeta'_L. \eta'_L + \BE ] - 
{t (t +2) \over (s-m^2)(u - m^2)} 
\eta_L.\eta'_L \, \zeta.Q_L \, \zeta'.Q_L
\big),
\\
A_1(s,t,u) = - \pi \G(-{t\over 2}) \G({m^2-u\over 2}+1)
\G({m^2-s\over 2}+1)/[\G({t\over 2} + 2) \G({u - m^2 \over 2})
\G({s - m^2 \over 2})]
\\
\ea
\label{1.5}
\ee
Here b.e. represent Bose exchange of particles 2 and 4.
\footnote{Our general notation for amplitudes is as follows.  
In the four-point scattering amplitude $\M(1,2,3,4)$ particles 1,3
refer to incoming and outgoing BH states and 2,4 refer to incoming and
outgoing states of the probe respectively. The polarization tensors of
the outgoing states are denoted by primes: thus the incoming
polariztions of BH's are $\zeta_{L,R}$ and outgoing polarizations are
$\zeta'_{L,R}$. Similarly those for the probes are $\eta_{L,R}$ and
$\eta'_{L,R}$ respectively. The charges of the particles 1,2,3,4 (when
they are non-zero) will be denoted by $Q_{L,R}, q_{L,R}, Q'_{L,R}$ and
$q'_{L,R}$ respectively. All momena and charges will be taken as
ingoing so that momentum conservation reads as $\sum_{i=1}^4 k_{i,\mu}
=0$ and charge conservation reads as $Q_L + Q'_L + q_L + q'_L = Q_R +
Q'_R + q_R + q'_R =0$. The Mandelstam varibles are defined as usual:
$s = - (k_1 + k_2)^2, t = -(k_1 + k_3)^2, u = - (k_1 + k_4)^2$.}

\gap{2}

\underbar{Determination of BH polarizations:}

\gap{2}

The various terms in \eq{1.5} involve {\em different} functions of the
Mandelstam variables and are hence independently measurable from
experiments. Of importance to us is the (post-Rutherford) term
\be
\ba
\M(1,2,3,4) = \ldots + \zeta_R.\eta_R \; \zeta'_R.\eta'_R \;
\zeta_L.\eta_L \; \zeta'_L.\eta'_L f(s,t,u) + \ldots, \\
f(s,t,u) \equiv \frac{(t/2)^2 (t/2 + 1)}{[(s-m^2)/2]^2 
[(s-m^2)/2 + 1]} A_1(s,t,u) = \sum_{n=0}^\infty {f_n(s) \over t 
- (2n + 2)},\\
f_0(s) =  -2 \pi (1 + (s-m^2)/2)
\ea
\label{1.6}
\ee
An experimental measurement of this term amounts to a measurement of
the combination $\zeta_R.\eta_R \; \zeta'_R.\eta'_R \; \zeta_L.\eta_L
\; \zeta'_L.\eta'_L f(s,t,u).$ It is therefore easy to see that by
repeating the experiment with various choices of $\eta_{L,R}$ we can
determine the initial polarizations $\zeta_{L,R}$ of the black hole.
The fact that $f(s,t,u)$ has no pole at $t=0$ implies that the
information about the polarizations $\zeta$ is propagated by stringy
modes. We will have more to say on this in the next section on stringy
classical backgrounds.

\gap{2}

\underbar{Black holes with $N_L > 1$:}

\gap{2}

The calculation for $N_L=1$ can be generalized in a straightforward
fashion to higher $N_L$. In this case there are higher tensors
$\zeta_L$ which describe polarization of the black hole,
e.g. $\zeta_{L, i_1i_2\ldots i_n}$. Once again there are
post-Rutherford terms, similar to \eq{1.7}. The general term involving
entanglement of polarizations of BH and probe is of the form
\be
\ba
\M(1,2,3,4) = \ldots + \zeta_R.\eta_R \; \zeta'_R.\eta'_R \;
\eta^i_L \,\zeta_{L, i j_1 j_2 \ldots j_n } \, \zeta_L^{\prime k j_1
j_2  \ldots j_n} \,\eta'_{L, k}\; g(s,t,u) + \ldots, \\
g(s,t,u) =  \sum_{n=0}^\infty {g_n(s) \over t - (2n + 2)}
\ea
\label{1.7}
\ee
Once again this term can be separated from the rest of the terms in
the $S$-matrix by its momentum dependence.  It is clear that by
choosing the polarization of the massless probe appropriately, we can
glean some information about the quantum state of the black hole. The
information is, however, partial since most of the ``legs'' of the
tensor $\zeta_L$ do not contract with $\eta_L$. In order to improve
this situation, we need to consider massive probes (neutral or
charged).

\gap{2}

(b)  Massive probes:

\gap{2}

Massive states can either be neutral or charged. A particle
with charge $q_{L,R}$ has a mass
\be
m^2 = q_R^2 +  2 (N_R - 1/2) = q_L^2 + 2 (N_L -1)
\label{1.8}
\ee

(i) \underbar{Neutral probes}:

Let us consider neutral probes first ($q_{L,R} =0,
N_R -1/2 = N_L -1 = n$).  Their masses are
of the string scale:
\be
m^2 =  2n
\label{1.9}
\ee
As in the case of the massless probes above (which are necessarly
neutral), the charge of the black hole is unaffected by scattering
with these probes: $Q_{L,R} + Q'_{L,R} = 0$, since $q_{L,R} = q'_{L,R}
=0$.

The new feature that arises in the 4-point $S$-matrix is that we start
obtaining terms with increasing number of contractions between the
polarizations of the probe and the black hole. For example, a massive
probe with polarizations $\eta_{R, i_0 i_1 \ldots  i_n},\;
\eta_{L, j_0 j_1 \ldots  j_n}$ has the following term in the
$S-matrix$ (we have considered BH vertex operators \eq{1.2} with all
$n_i =1$):
\be
\ba
\M(1,2,3,4) = \ldots + 
\zeta_R^{i_0} Q_R^{i_1} \ldots Q_R^{i_n} \, \eta_{R,i_0 i_1 \ldots i_n} \,
\zeta_R^{\prime j_0} Q_R^{j_1} \ldots Q_R^{j_n} 
\, \eta'_{R,j_0 j_1 \ldots j_n} \\
\,~~\eta^{k_0 k_1 \ldots k_n}_L\,
\zeta_{L, k_0 k_1 \ldots k_n p_1 p_2 \ldots p_{N_L - n -1} } 
\zeta_L^{\prime l_0 l_1 \ldots l_n p_1
p_2 \ldots p_{N_L - n -1}} \, \eta'_{L, l_0 l_1 \ldots l_n} h(s,t,u)
+ \ldots,\\
\, h(s,t,u) = \sum_{m=0}^\infty h_m(s)/(t - (2m + 4n +2 ))
\ea
\label{1.10}
\ee
A measurement of this term for various choices of the probe
polarizations $\eta_{L,R}$ ultimately {\em determines} for us
$\zeta_{L,R}$ in case $n=N_L$, that is, when the (left) oscillator
level of the probe matches the (left) oscillator level of the black
hole. In the next section (on classical backgrounds) we will see that
this corresponds to a $\zeta$-dependent non-zero background value of a
string mode (mass$^2= 2n$) around the black hole. We should note here
that such a background can be detected either by a single massive
probe (as described by a four-point function) or by many massless
probes (which involves an $n$-point amplitude for large $n$).

\gap{2}

(ii)  \underbar{Charged Probes:}

\gap{2}

(1)  Kaluza-Klein probes:

\gap{2}

Let us first consider probes whose 10-dimensional masses vanish,
in other words, probes with $N_R - 1/2 = N_L -1 = 0$. Their
(four-dimensional) masses are 
\be
m^2 = \q_R^2 = \q_L^2
\label{1.11}
\ee
which are of the order of the compactification scale. We shall call
these probes KK probes.

With charged probes we can have two kinds of amplitudes, one
in which the charge does not change (neutral channel) and another in
which the charge changes (charged channel):

\gap{2}

\underbar{Neutral channel} 

\gap{2}

This is the case in which the charge of the probe does not change:
$\q_{L,R} = -\q'_{L,R}$. The charge of the black hole also remains the
same. Vis-a-vis the polarization tensors $\zeta$, the $S$-matrix
in this case does not give any more information than in the
massless case discussed above.

\gap{2}

\underbar{Charged channels}

\gap{2}

These are amplitudes in which the charge of the black hole and the
charge of the probe are allowed to change:
\be
Q_{L,R} + Q'_{L,R} =  \q_{L,R} + \q'_{L,R} = \Delta \q_{L,R} \ne 0
\label{1.12}
\ee
This also typically implies a change of $N_L$. We will call the final
oscillator level of the black hole $N'_L$. Note that this is an
inelastic process, that is, the BH under measurement changes its state
after the ``measurement''. However, by measuring such amplitudes, we
can still get useful information about the initial state of the black
hole. Indeed, $S$-matrix elements of this type can completely
determine the initial polarization tensor $\zeta$ of the black
hole. It can be shown that the $S$-matrix contains a term (which can
be independently measured by choosing the momenta and charges of
initial and final states of the probe appropriately)
\be
\ba
\M(1,2,3,4) = 
\ldots +
\zeta_R. \eta_R \; \zeta'_R. \eta'_R \\
\; \Delta \q_{i_1} \Delta \q_{i_2} \ldots \Delta \q_{i_{N_L}} 
\zeta_L^{i_1 i_2 \ldots i_{N_L}} \;
\Delta \q_{j_1} \Delta \q_{j_2} \ldots \Delta \q_{j_{N'_L}} 
\zeta_L^{\prime j_1 j_2 \ldots j_{N'_L}} 
F(s,t,u)\\
\,~ F(s,t,u) = \sum_{n=0}^\infty {F_n(s) \over  t - (m_{KK}^2 + 2n)}
\\
\,~ m_{KK}^2 \equiv \Delta \q_R. \Delta \q_R = \Delta \q_L . \Delta q_L
\ea
\label{1.13}
\ee
The second equality in the last line follows from the fact that we are
considering exchange particles satisfying $N_R - 1/2 = N_L -1 =0$.

\gap{2}

\underbar{Determination of hair:}

\gap{2}

In the above $\Delta \q_L \equiv \q_L - (- \q'_L) $ is the charge
difference between the initial and final states of the probe and is
therefore a known vector.  By a sufficient number of experiments with
various choices of $\Delta \q_L$ we can ultimately determine the
tensor $\zeta_L$. The determination of $\zeta_R$ is trivial (by
tailoring $\eta_R$). Thus we are able to determine the initial state
of the black hole by using Kaluza-Klein probes whose generation
requires energies of the compactification scale rather than the string
scale.  Such a determination of the internal state of the black hole
bears a close resemblance to the observations made in
\cite{larwil}. We will have more to say on this at the end of the next
section.

\gap{2}

(2) Charged probes of string mass: These do not contain any new
physics and so we do not consider them here.

\gap{3}

\section{Hairy classical backgrounds} 

In this section we show how to interpret the earlier results as the
scattering of probe particles off backgrounds of string modes,
massless {\em as well as massive}, created by the BPS state.  Let us
denote the string mode in spacetime corresponding to the BPS state by
$\psi(x)$ (this field has indices: $\psi^{i_1 i_2\ldots i_n}$
corresponding to polarization indices in the internal compact
directions). The beta function equations for neutral fields, say
$\phi^{k_1\ldots k_m \, \mu_1 \ldots \mu_l}$ are of the form
\cite{polyakov,klesus}
\be
\ba
0= \beta_\phi^{k_1 \ldots k_m\, \mu_1 \ldots \mu_l}(q) = (q^2 +
m_\phi^2) \phi^{k_1 \ldots k_m\, \mu_1 \ldots \mu_l}(q) 
\\ 
\, 
- g_{st} \int \! d k_1 d k_3 [\Gamma_\phi^{\psi\psi}
]^{k_1 \ldots k_m \, \mu_1 \ldots \mu_l}_{i_1 \ldots i_n, \, j_1
\ldots j_n} (k_1, k_3) \psi^{* \, i_1 \ldots i_n}(k_1) \psi^{j_1
\ldots j_n} (k_3)
\tilde \delta^4 (q - k_1 - k_3 ) + \cdots \\
\ea
\label{2.1}
\ee
which we will schematically write as (suppressing indices)
\be
0 = (q^2 + m_\phi^2) \phi(q) - g_{st}
\int d k_1 d k_3 \tilde \delta^4 ( q - k_1 -
k_3) \Gamma_\phi^{\psi \psi} (k_1 , k_3) \psi^*(k_1) \psi(k_3)
+ \ldots
\label{2.2}
\ee
Here $\Gamma^{\psi\psi}_\phi$ are determined by the operator product
expansion coefficients \cite{polyakov,klesus} of the vertex operetors
of the fields $\psi, \psi$ and $\phi$. Also, we have used the 
notation $dk \equiv d^4k/(2 \pi)^4$ and $\tilde \delta^4 (k)
\equiv (2 \pi)^4  \delta^4(k)$. 

We are interested in the BPS state, described by the field $\psi$ (mass
$m$), to be in its rest frame and given by a wavefunction
\be 
\psi^{(0), i_1 \ldots i_n} (k) = 2 \pi \delta(k_0 -
\omega_{\vec k}) \zeta^{i_1 \ldots i_n} f(\vec k) / 
\sqrt{2 \omega_{\vec k}}, 
\quad   \omega_{\vec k}^2 \equiv  {\vec k}^2 + m^2,
\label{2.3} 
\ee 
where $f(\vec k)$ is a wave-packet centered around $\vec k =0$ and
satisfies $ \int {d^3\vec k \over (2\pi)^3} | f(\vec k) |^2 = \int d^3
x \, \psi^* i {\buildrel \leftrightarrow \over \del_0} \psi
=1$. $\zeta^{i_1 \ldots i_n}$ denotes the (real) polarization tensor
of the particle and satisfies the normalization condition $\zeta^{i_1
\ldots i_n}
\zeta_{i_1 \ldots i_n} = 1$.  The $\phi$
background created by such a BPS state is obtained, to first order in
$g_{st}$, by solving \eq{2.2}:
\be
\phi^{(1)} (q) = g_{st} (q^2 + m_\phi^2)^{-1} 
\int \! d k_1 d k_3 \Gamma^{\psi \psi}_\phi 
(k_1, k_3, q) \psi^{* (0)} (k_1) \psi^{(0)} (k_3) \delta( k_1 + k_3 -
q)
\label{2.4}
\ee
It is not difficult to show that both \eq{2.3} and \eq{2.4} are
classical solutions of their respective equations of motion to first
order in $g_{st}$. In the case of sufficiently peaked wave-packets
$f(\vec k)$, we can easily carry out the integral \eq{2.4}, after
substituting for $\psi^{(0)}$ from \eq{2.3}. We get
\be
\phi^{(1), k_1 \ldots k_m, \mu_1 \ldots \mu_l} (q) = 
{g_{st} \over 2m } (q^2 + m_\phi^2)^{-1} [\Gamma^{\psi \psi}_\phi
]^{k_1 \ldots k_m, \mu_1 \ldots \mu_l}_{i_1 \ldots i_n, j_1 \ldots
j_n} (\bar k, \bar k, q) \zeta^{i_1 \ldots i_n} \zeta^{j_1 
\ldots j_n} 2 \pi \delta(q_0)
\label{2.5}
\ee
where $\bar k \equiv (m, \vec 0)$ denotes the four-momentum of the
BPS state in its rest frame. 

In deriving the above solution we have used the condition of low
momentum transfer ($| \vec q| \ll m_{string}$). While we expect the
solution to get modified at short distances due to string world sheet
corrections \cite{sen}, it should remain valid at large distances.  We
should also mention that \eq{2.5}, together with \eq{2.3}, represent
classical backgrounds corresponding to an {\em infinite} number of
string modes, and it is the set of all these backgrounds which
constitutes the string theory black hole.

\gap{2}

\underbar{Scattering off background:}

\gap{2}

Let us now find the scattering amplitude of a probe particle
(represented by some string mode $\chi(x)$ of mass $\mu$ say) off the
above background \eq{2.5}. The relevant part of the action is
\be
\ba
S = \int dk\,  (k^2 + \mu^2) \chi(k)\chi(-k) + 
\\
\, g_{st}\int dk_2 dk_4 dq\,
\Gamma^{\chi \chi}_\phi \chi(k_2) \chi(k_4) \phi(q) \delta (k_2
+ k_4 + q) + \ldots
\ea
\label{2.6}
\ee
The S-matrix describing the amplitude for scattering of the field
$\chi(x)$ ($k_2, k_4$ are the initial and final momenta and $\eta,
\eta'$ are the initial and final polarizations) is given by
\be
{\cal S} = \frac{\phi^{(1)} \Gamma^{\chi \chi}_\phi \eta \eta' }
{\sqrt{2 k^0_2}\sqrt{ 2 k^0_4}}
\label{2.7}
\ee
where we have dropped the momentum conserving delta-functions.
If we substitute for $\phi^{(1)}$ from \eq{2.4}, we get
\be
{\cal S} = \frac{\Gamma^{\psi \psi}_{\phi} 
\zeta \zeta  \Gamma^{\chi \chi}_\phi  \eta \eta'}{
(k_2 + k_4)^2 + m_\phi^2 } \frac{1}{ \sqrt{2 k^0_1}\sqrt{2 k^0_2}
\sqrt{2 k^0_3}\sqrt{2 k^0_4}}
\label{2.8}
\ee
where $k^0_1 = k^0_3 = m$. 

\gap{2}

\underbar{Comparison with 4-point amplitude}:

\gap{2}

It is easy to show that \eq{2.8} is precisely the $S$-matrix for the
process in which the particle $\chi(x)$ with initial momentum $k_2$
and initial polarization $\eta$ is scattered to a state $(k_4, \eta')$
because of a $\phi$-particle exchange with a static BPS state with
mass $m$ and polarization $\zeta$. We see, therefore, that the 4-point
$S$-matrices calculated in the previous section, factorized on
specific channels, can be interpreted as scattering off classical
backgrounds given by expressions like \eq{2.5}. The important thing
to note is that these expressions explicitly involve the polarization
tensor $\zeta$ of the black hole. 
\gap{2}

\underbar{Determination of the backgrounds}

\gap{2}

The strategy for determining the (first order) backgrounds around the
BPS state is now clear. We either use \eq{2.5} directly, or solve
\eq{2.7} for $\phi^{(1)}$ from a knowledge of ${\cal S}$, the
4-point $S$-matrix. As we have shown above, the two procedures
are equivalent.

We now list some backgrounds which we determine using this method.
For massless exchanges, we get \cite{garmey}
\be
h^{(1)}_{\mu \nu}(\rho) = m/\rho \delta_{\mu 0} \delta_{\nu 0}, \quad
\Phi^{(1)} (\rho) = -m / (2 \rho)
\label{2.12}
\ee
These agree with the first order backgrounds of metric and dilaton as
given in \cite{sen}. For the metric this agreement has already been
shown in \cite{manwad}. 

\gap{2}

Massive backgrounds:

\gap{2}

The amplitude \eq{1.6} does not have a massless $t$ channel. The
lightest exchange particles are of mass $m^2=2$.  We list below the
background value of one of these, corresponding to vertex operator
$V^{ij} \bar\del x^k \bar\del x^l$ \footnote{
$V^{ij} = \del x^i \del x^j + \del \psi^i \psi^j + i (k.
\psi) \psi^i \del x^j$ is the supersymmetrized version
of $\del x^i \del x^j$.}  (again determined using the
method outlined above) 
\be
M^{(1)}_{ijkl}(\rho) = (1/m) e^{- \sqrt{2} \rho}/\rho \; [\zeta_{R,i}
\zeta_{R,j} +  Q_{R,i}Q_{R,j}]
[\zeta_{L, k}\zeta_{L,l} + 1/2 Q_{L,k}Q_{L,l}] + o(g_{st})
\label{2.13}
\ee
Thus we have backgrounds which explicitly carry hair. 

In case of higher $N_L$ black holes, the above string mode has a
background value (corresponding to \eq{1.7})
\be
M^{(1)}_{ijkl}(\rho) = (1/m)  e^{-\sqrt{2} \rho}/\rho 
[\zeta_{R, i} \zeta_{R,j}
\zeta_{L,k j_1 \ldots j_n} \zeta_{L,l}^{j_1 \ldots j_n}
+ \ldots] 
 + o(g_{st})
\label{2.14}
\ee
In the above, the ellipsis represents additional terms which depend on
the polarization and the charge vectors. These terms appear in the
$S$-matrix elements separately. Clearly, because of the internal
contractions within the $\zeta$'s these bckgrounds do not carry enough
hair to determine the $\zeta$'s completely.

\gap{2}

More massive backgrounds:

\gap{2}

If we consider backgrounds of sufficiently heavy string modes, they
carry enough ``hair'' so as to let us determine $\zeta_L$ completely.
In the context of four-point amplitudes these backgrounds are
seen only by massive string modes. Thus the background corresponding
to the lightest channel of the amplitude \eq{1.10} is given by 
\be
\ba
M^{(1), i_0 i_1 \ldots i_n, j_0 j_1 \ldots j_n, k_0 k_1 \ldots k_n,
l_0 l_1 \ldots l_n}\\
=(1/m) [e^{-\sqrt{4n + 2}\; \rho}/ \rho]
\zeta_R^{i_0} Q_R^{i_1} \ldots Q_R^{i_n} \, 
\zeta_R^{\prime j_0} Q_R^{j_1} \ldots Q_R^{j_n} 
\\
\zeta^{k_0 k_1 \ldots k_n}_{L,p_1 p_2 \ldots p_{N_L - n -1} } 
\zeta_L^{\prime l_0 l_1 \ldots l_n p_1
p_2 \ldots p_{N_L - n -1}} 
\ea
\label{2.15}
\ee
Clearly for $n$ sufficiently large, there will be no internal
contractions between the $\zeta$'s and therefore by measuring
sufficiently heavy backgrounds we can see enough ``hair'' so as to
determine the polarization tensors $\zeta$.

\gap{2}

\underbar{Interpretation of scattering involving charged channels:}

\gap{2}

The background fields determined by the above procedure correspond to
neutral excitations of the string. As we have seen, however, (in the
subsection containing equations \eq{1.12} and \eq{1.13}), inelastic
amplitudes corresponding to a charge exchange between the BH and the
probe particles, are also useful in determining the initial state of
the black hole, and in fact, can be used to completely fix the initial
state. The KK backgrounds (hair) in the classical solution in
\cite{larwil} seem to correspond to such charge exchanges between the
BH and the probe particles. In this connection we should note that
\cite{calmalpee} identifies the collective string Hilbert space,
obtained by quantizing the functions $\vec{f_a}(u), p_a(u)$ and
$q^I_{L,a}(u)$ (see Sec. 2.2 of \cite{calmalpee}) appearing in the
classical solution, to be the Hilbert space of the elementary string.

We can, in fact, make a more precise connection of the scattering
amplitudes involving charged channels with the long-range KK `hair'
emphasized in \cite{larwil}. At a first sight it would seem that the
scattering in the charged channels is suppressed, at low momentum
transfer, by a factor of $1/m^2_{KK}$ because of the propagator of the
KK particle, $1/(t - m^2_{KK})$. However, if we consider processes in
which the final state of the black hole is also a BPS state, with a
charge vector $Q_R$ which is collinear (but not identical) with
$Q'_R$\footnote{We note that the KK backgrounds considered in
\cite{larwil} correspond to momentum in the $x^9$ direction, hence the
charge vectors are collinear in an obvious sense; similarly our
requirement of the final state satisfying the BPS condition is similar
to their requirement that the background be supersymmetric.}, then, at
low momentum transfer, the kinematics of the process `conspires' to
remove the $1/m^2_{KK}$ suppression! To see this, note that the
4-momentum carried by the exchanged particle (in the rest frame of the
initial BPS state) is $q_\mu = (q_0, \vec q)$ where
\be
q_0 = \sqrt{ m^{\prime 2} + | \vec q |^2 } - m,
\quad 
\vec q = -(\vec k_1 + \vec k_3),
\label{2.16}
\ee
where $m = | Q_R |$ and $m' = | Q'_R |$ are the masses of the initial
and the final BPS state. In the limit $|\vec q | \ll m'$, we have $
q_0 \approx m'- m$. Now, the mass of the exchanged KK particle is given
by $m^2_{KK} = | Q'_R - Q_R |^2$. For collinear charges $Q'_R$ and
$Q_R$, this is exactly equal to $ (m' -m)^2$. We, therefore, find that
for such processes the propagator for the exchanged KK particle goes
as
\be
\frac{1}{t - m^2_{KK}} = - \frac{1}{ | \vec q |^2}
\label{2.17}
\ee
which is characteristic of a long-range interaction! It is important
to note here that for such processes the term in the charge exchange
amplitude given in \eq{1.13} is non-zero only for those black holes
which carry charges in the sixteen toroidal directions of the
uncompactified heterotic string and have polarizations in these
directions as well as in the KK directions. 

The charge exchange scattering processes identified above are
obviously not generic enough to determine the internal polarization
tensor of the black hole completely. To do this, we need to consider
the more general processes in which $Q_R$ and $Q'_R$ are not
collinear. In this case, the `miracle' in \eq{2.17} does not happen,
and we have a suppression of these amplitudes by $1/ m_{KK}^2
$. Nevertheless, it is interesting that the mass scale at which
information about the initial polarization tensor of the black hole
can be obtained from scattering experiments is $m_{KK}$ and not
$m_{string}$.  Such a conclusion has been anticipated in
\cite{larwil,calmalpee}.

\gap{3}

\section{D-branes} 

So far we have discussed the case of electrically charged black holes.
It would clearly be interesting to see how the above ideas apply to
the case of fat black holes described by D-branes \cite{fat}. This
work is in progress, but we have some preliminary results for single
D-branes carrying single open string excitations. We will show below
that amplitudes involving the scattering of closed strings off such
excited D-branes encode information about the polarization of the open
string excitations.

We consider type II superstring in $R^{d-1,1} \times T^p, d + p =10$
and a $Dp$-brane \cite{pol} wrapped on the $T^p$.  Our notation for
the spacetime coordinates is $ x^M = (x^\mu, x^i), \mu = 0, 1, \ldots,
d-1; \; i= 1,\ldots, p$. We will also use the notation $x^a = (x^0,
x^i)$ for directions parallel to the D-brane, which include time.  We
consider a single open string excitation on the $D$-brane with
polarization $\zeta^a = (\zeta^0, \zeta^i)$ which is parallel to the
D-brane. The vertex operator of this excitation is
\be
V_{E}(\zeta^a,p^a,z) = \zeta_{a} (\partial x^{a} +ip_{b}\psi^{b}
\psi^{a})\exp (ip.x(z))
\label{3.1}
\ee
Note that this excitation moves along the compactified directions, so
its momenta $p^i$ are quantized. We now consider scattering the
following closed string probes off the excited D-brane:
\be
\ba
\V_{P}(\eta_{R},\eta_{L};k^\mu ;\q^i;z,\bar{z}) = \\
\, ~~~ V_{P}(\eta_{R},k,\q ,z)
\bar{V}_{P}(\eta_{L},k, \q, \z)\exp[ik\mu x^\mu (z,\z)+i\q_i x^i (\z)], 
\\
V_P(\eta_R,k,\q,z)=\eta_{R,M}(\partial x^{M} + (i k_\mu
\psi^{\mu}(z) + i \q_i \psi^i(z) ) \psi^{M}(z)), \\
\bar V_P(\eta_L,k,\q,z)=\eta_{L,M}(\bar \partial x^{M} + (i k_\mu
\psi^{\mu}(\z) + i \q_i \psi^i(\z) ) \psi^{M}(\z))
\ea
\label{3.2}
\ee
where $k$ is the space time momentum and $\q$ is the KK charge.  We
consider for simplicity probes with no winding modes: $\q_R = \q_L$;
these turn out to be sufficient for our purposes here.  The connected
amplitude for the process at tree level is a disc diagram with two
open strings at the boundary and two closed strings in the interior.
It is easy to compute the part of this amplitude which arises from
the exchange of closed strings which are massless in the ten-dimensional
sense. One such term in the $S$-matrix (which can be separately
measured by choosing the probe momenta and charges appropriately) is
reproduced below
\be
\ba
\M(1,2,3,4) = \ldots + 
[16 \pi q_{\parallel}^4 \Gamma (2q_{\parallel}^2)/(\Gamma (1+
q_{\parallel}^{2}))^2 ( q^2 + m_{KK}^2 )] \times
\\
\left( \zeta'. \eta_R \; (\eta_{L, \mu}q^\mu + \eta_{L, i} \Delta \q^i)
(\eta'_{R, \mu}q^\mu + \eta'_{R, i} \Delta \q^i)
\; \zeta . \eta'_L + \BE \right) \times \\
\left( \zeta'.\eta_L \; (\eta_{R, \mu}q^\mu + \eta_{R, i} \Delta \q^i)
(\eta'_{L, \mu}q^\mu + \eta'_{L, i} \Delta \q^i)
\; \zeta .\eta'_R + \BE \right) + \ldots \\
\ea
\label{3.3}
\ee
In the above $q = k_2 + k_4, \; q_{\parallel} = -q_{0}^{2} + \vec q^2$
and $\Delta \q (\equiv \q + \q')$ denotes the charge difference between
the initial and final states of the probe. It is clear that by choosing
probes appropriately we can determine the polarization $\zeta^a$. 

We should note that the above amplitude corresponds to the exchange of
a charged closed string (KK particle). It can be shown that neutral
massless channels do not exhibit any `hair', that is, they do not have
any entanglement between the polarizations of the probe and the open
string excitation on the D-brane.

\gap{2}

\underbar{Hair in absorption and decay amplitudes of fat black holes:}

\gap{2}

The above calculation addresses a BPS black hole made up of 
only a single D-brane and open string excitations on it. The 
construction of fat black holes \cite{fat}, of course, involves 
multiple D-branes and open string excitations between them.
It is an interesting question in that case whether one needs open
string probes \footnote{We thank F. Larsen and F. Wilczek for raising
this issue.}  to completely determine the states of a black hole,
since the latter form a non-trivial representation of a $U(N)$ gauge
theory where closed string probes are singlets under such gauge
groups. Work on this problem is in progress and we hope to come back
to this question.  Meanwhile, we close this section with a brief remark
about an $S$-matrix calculation  presented in \cite{dhamanwad}
which exposes more detail about the quantum state of a fat black hole
than is warranted by the classical no-hair theorems. These black holes
are represented by left- and right-moving open string excitations on
D-branes (for more detail, we refer the reader to \cite{dhamanwad} and
references therein) and their detailed quantum state is represented by
eqn. (6) of \cite{dhamanwad} which involves the number distribution of
open strings with specific left- and right-moving momenta. The
$S$-matrix element involved in the absorption of a closed string
quantum by the black hole (same as the $S$-matrix for the decay of
closed strings from the black hole) is given by eqn. (8) of
\cite{dhamanwad}. If the closed string quantum coming out in the decay
process has an energy $\omega$ integer, then the $S$-matrix actually
involves the number of left- and right-moving open strings with
momentum $\omega/2$ (denoted as $N_{L,R}(m)$ in eqn. (8) of
\cite{dhamanwad}) and the latter can therefore be measured from
$S$-matrix data. Note that it is only the microcanonical average of
these number distributions (and not the individual number for each
state) which is determined by the temperature of the black hole.  The
individual distributions of left- and right-movers obviously have much
more information than is contained in the the data allowed by no-hair
theorems.

\gap{3}

\section{Conclusion}

In this paper we have presented a computation of $S$-matrix for
scattering of probe particles off a black hole and shown how a
measurement of the post-Rutherford terms can uniquely determine the
microstate of the black hole from the $S$-matrix data. The calculation
has primarily been carried out for the electrically charged black
holes of \cite{sen} and some preliminary results have been presented
for D-branes.  The fact that we are able to determine the state of the
black hole from an $S$-matrix is consistent with the identification of
the black hole state with an elementary string state; however, it
immediately raises the interesting question of what happened to the
usual no-hair theorems of general relativity, which seem to preclude
such detailed measurement of the state of the black hole from
outside. This question becomes particularly intriguing in the light of
our observation (Sec. 2) that some of the measurements only require
energies of the order of the compactification scale which can be far
less than the string scale. Our statements about the $S$-matrix appear
to be closely connected to the observations made in \cite{larwil}
about violation of no-hair theorems in the context of classical
solutions. Indeed, we explicitly demonstrate the interpretation of
some of our $S$-matrix elements in terms of stringy classical
backgrounds which carry information about the detailed state of the
black hole. We would like to remark that in regimes where differential
equations satisfied by various string modes cannot be trusted, the
$S$-matrix may provide a more operational {\em definition} of various
backgrounds (this is similar to the operational definition of horizon
area proposed in \cite{sus} as the absorption cross-section).  Such an
$S$-matrix approach to a consistent unitary quantum mechanics for
black holes has actually been advocated as a principle by 't Hooft
\cite{thooft}.  According to such a philosophy, it is the emergence of
the classical no-hair theorems from a unitary $S$-matrix which
requires an explanation. What we have seen in the present work is that
the terms in the amplitude that exhibit the no-hair property are
associated with massless exchanges; in the presence of string modes or
Kaluza-Klein modes this property is lost, thus enabling determination
of the state of a black hole, much like in the case of ordinary
matter.

\vspace{3 ex}

Acknowledgement: We would like to thank F. Larsen, Y. Nambu, A. Sen,
L. Susskind, G. Veneziano and F. Wilczek for critical discussions.
G.M. would like to acknowledge the hospitality of Princeton
University, the Institute for Advanced Study, Princeton, Caltech and
Strings '96, Santa Barbara, where part of the work was done.

\end{document}